\documentclass{article}
\usepackage{amsmath}
\usepackage{graphicx} 
\usepackage[dvipsnames]{xcolor}
\usepackage{authblk}
\usepackage{hyperref}

\title{Wobbling and Migrating Ferrofluid Droplets}
\author[1]{Aaveg Aggarwal*}
\author[2]{Shih-Yuan Chen*}
\author[1]{Eleftherios Kirkinis}
\author[4]{Mohammed Imran Khan}
\author[4]{Bei Fan}
\author[2]{Michelle M Driscoll}
\author[$\dagger$1,2,3,5]{Monica Olvera de la Cruz}
\affil[1]{Department of Materials Science and Engineering, Northwestern University, Evanston, IL, USA}
\affil[2]{Department of Physics and Astronomy, Northwestern University, Evanston, IL, USA}
\affil[3]{Department of Chemistry, Northwestern University, Evanston, IL, USA}
\affil[4]{Department of Mechanical Engineering, Michigan State University, East Lansing, MI, USA}
\affil[5]{Department of Chemical and Biological Engineering, Northwestern University, Evanston, IL, USA}
\affil[$^\dagger$]{Corresponding author: m-olvera@northwestern.edu}

\date{}
\begin{document}

\maketitle

\maketitle
\begin{abstract}
Active components incorporated in materials generate motion by inducing conformational changes in response to external fields. Magnetic fields are particularly interesting as they can actuate materials remotely.  Millimeter-sized ferrofluid droplets placed on a solid surface, surrounded by an ambient gas phase, are shown here to migrate under a rotating magnetic field due to the periodic deformation of the liquid-gas interface. This interface wobbling leads to droplet migration with speeds that increase as the amplitude and frequency of the magnetic field increase. In addition to migrating in a controlled manner, we demonstrate the ability of magnetic droplets to clean surface impurities and transport cargo. 
\end{abstract}

\section{Introduction}
Soft active materials are characterized by their dynamic response to external stimuli. Unlike passive materials, active materials have the ability to respond to different inputs such as light \cite{aggarwal2022controlling, zhao2019soft, palacci2013photoactivated}, magnetic fields \cite{driscoll2019leveraging, jackson2017ionic}, electric fields \cite{zhang2022polar, pradillo2019quincke}, and chemical cues \cite{li2021chemically, maeda2007self, manna2023chemically}. Active components incorporated in passive materials enable controlled conformational changes to achieve specific functions such as migration, swimming, and delivering cargo. Moreover, soft materials actuated by magnetic fields hold significant appeal due to the fact that magnetic fields can penetrate a wide range of materials, including biological matter. 

Ferrofluids are colloidal suspensions of magnetic nanoparticles that can be actuated using external magnetic fields. Since the motion of magnetic particles inside the host fluid can generate macroscopic fluid flow, the magnetic fields enable tunable fluid control in situ without changing the fluid properties and confinement. 
Different actuation schemes can be devised to manipulate magnetic liquids. Ferrofluids and other magnetic materials have been explored to achieve locomotion via spatial gradients in magnetic fields. These gradients can be established by using magnets \cite{sarkhosh_manipulation_2023} or current carrying coils \cite{nguyen_kinematics_2007}. The magnetic field source can also be physically moved \cite{nguyen_magnetowetting_2010} or periodically turned off and on \cite{fan_reconfigurable_2020}, to create fields with spatio-temporal variations. 

Rotating magnetic fields have also been used to actuate matter for robotic applications \cite{li2020fast, li2023magnetic}. For ferrofluid droplets, this approach relies on the motion of the fluid for actuation. Although the individual magnetic constituents of the ferrofluid experience no net force in a spatially uniform external field, the magnetic particles rotate to align with the field and drag the surrounding fluid along, causing macroscopic fluid motion \cite{Kirkinis2017, aggarwal2023activity}. 
Rotational fields have been used to drive ferrofluid droplets in liquid environment \cite{sun_exploiting_2022}, pattern and control the moving direction of a pack of ferrofluid droplets \cite{fan_ferrofluid_2020}, and to direct ferrofluid droplets along magnetic rails \cite{katsikis_synchronous_2015}. High frequency rotating fields can create internal torques in ferrofluids causing the fluid to rotate along with the field \cite{chaves2008spin, rosensweig1990magnetic, rosensweig2013ferrohydrodynamics}. This phenomenon can also be used to displace ferrofluid droplets on solid substrates \cite{aggarwal2023activity} as the droplet fluid develops internal rotations. 

In this work, we show numerically and verify experimentally, that in the regime of negligible torques (no internal rotations), ferrofluid droplets on a solid substrate surrounded by an ambient gas phase, can migrate by applying an external rotating magnetic field. The liquid-gas interface deforms in the direction of the magnetic field \cite{rosensweig2013ferrohydrodynamics}, whose circulation causes the droplet and its contact lines to wobble. We find that this geometric wobbling of the droplet and its interaction with the solid substrate, causes the droplets to migrate. We develop a finite element analysis model to study the magnetic deformation and motion of these droplets. We also present experimental results which demonstrate the 
same wobble and migration effect existing in a ferrofluid droplet in the presence of a rotating magnetic field, thus verifying the overall trend of the numerical predictions.

\begin{figure}[!ht]
    \centering
    \includegraphics[width = \linewidth]{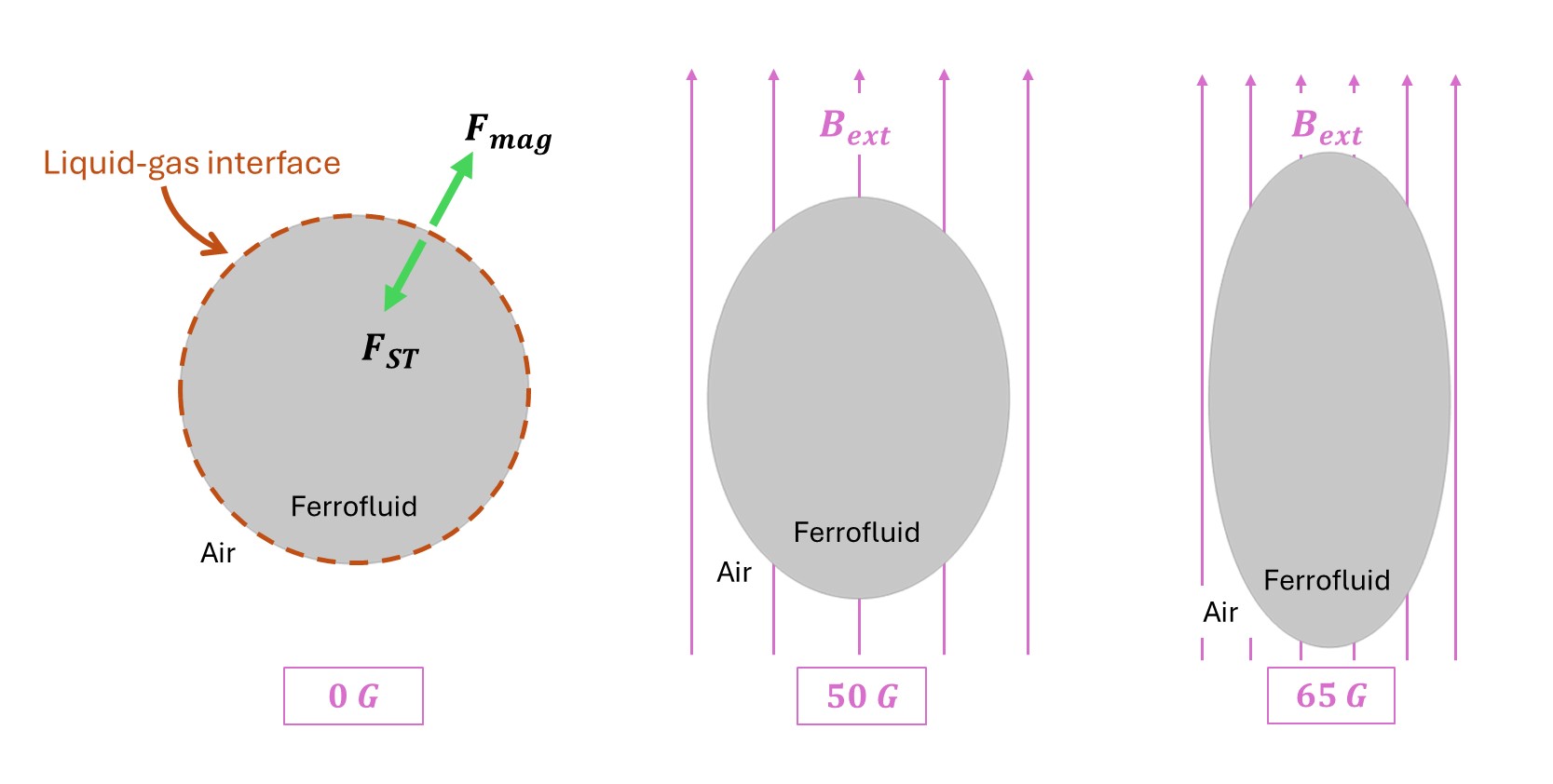}
    \caption{Finite element simulation of a ferrofluid droplet suspended in air in the presence of three different external magnetic fields $B_{ext}$. The geometry of the liquid-gas interface is determined by the balance of magnetic force $F_{mag}$ and the force due to surface tension $F_{ST}$. }
    \label{fig:1}
\end{figure}

\section{Results and Discussion}
Ferrofluids are incompressible liquids endowed with the 
magnetic stress tensor \cite{rosensweig2013ferrohydrodynamics}
\begin{equation}
    \label{eq:magn_stress}
    \boldsymbol{\sigma_m} = -\left\{ \mu_0 \int_0^H M dH + \frac{1}{2}\mu_0 H^2 \right\} \boldsymbol{I} + \boldsymbol{B} \boldsymbol{H}.
\end{equation}

Here, $\boldsymbol{M}$ is the fluid magnetization, $\boldsymbol{H}$ is the magnetic field, and $\boldsymbol{B}$ is the magnetic flux density. We use bold symbols to denote vectors and tensors, and non-bold symbols for their respective magnitudes. Therefore, the ferrofluid experiences a magnetic force $\boldsymbol{f}$ ($ = \textrm{div}~\boldsymbol{\sigma_m}$) given by.

\begin{equation}
    \boldsymbol{f} = -\mu_0  \boldsymbol{\nabla} \int_0^H  M dH  + \mu_0 M \boldsymbol{\nabla} H.
\end{equation}

When the magnetization relaxation time is very small the 
magnetization is collinear with the magnetic field at all times. If in addition the system is isothermal, the magnetization only depends on the field $\boldsymbol{H}$, leading the magnetic body force $\boldsymbol{f}$ 
to vanish \cite{rosensweig2013ferrohydrodynamics}. However, the magnetic body force entering into the interfacial force balance is \emph{non-zero}, causing a commensurate geometric deformation of the liquid-gas interface. The magnetic interfacial force is given by 

\begin{equation}
    \label{eq:interfacial_force}
    [[\boldsymbol{n}\cdot \boldsymbol{\sigma_m}\cdot \boldsymbol{n} ]]= \mu_0 \int_0^H M dH + \frac{1}{2}\mu_0 (\boldsymbol{M}\cdot \boldsymbol{n})^2,
\end{equation}
where $[[\cdot]]$ denotes the jump in the field across the interface of the droplet, and $\boldsymbol{n}$ is the unit vector normal to the droplet's interface. Fig. \ref{fig:1} shows a finite element simulation that demonstrates the effect of this interfacial force using a ferrofluid droplet suspended  in air. The interface of the droplet, suspended in air, becomes elongated along the direction of the magnetic field, while at the same time it is stabilized by surface tension.

\begin{figure}[!ht]
    \centering
    \includegraphics[width = \linewidth]{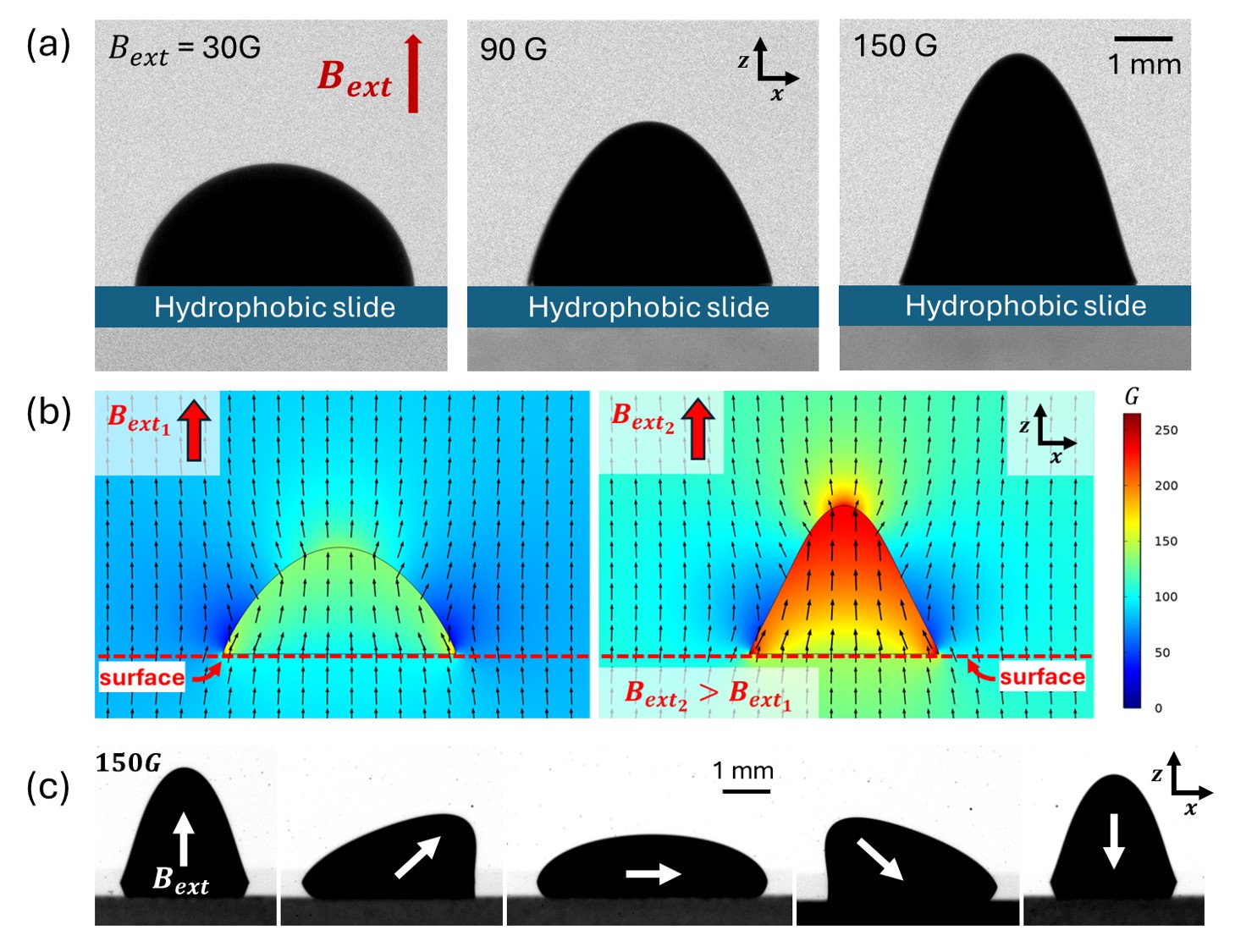}
    \caption{Deformation of the droplet interface due to magnetic field. (a) Photographs from experiments demonstrating an increase in the peak height of the ferrofluid droplet with an increase in the external field strength, $B_{ext}$. Magnetic fields used here are $30$ G, $90$ G and $150$ G and, the field points in the $+z$ direction. (b) Finite element model simulating the deformation of a 2D ferrofluid droplet as a function of increasing magnetic field strength. (c) Chronologically arranged photographs from left to right showing the droplet deformation during a half-cycle of clockwise rotating magnetic field. White arrows indicate the direction of the external field. }
    \label{fig:2}
\end{figure}

A ferrofluid droplet placed on a solid substrate also experiences the same magnetic force and undergoes a similar elongation. 
To test this, we use a commercial water based ferrofluid (FerroTec, EMG 700) and deposit a droplet of the fluid with $\sim 1.5$ mm radius on a chemically-coated hydrophobic glass slide (see Methods section~\ref{sec:method_exp} for more details). The droplet is then exposed to a magnetic field pointing in the $+z$ direction. Fig. \ref{fig:2}a shows the droplet geometries in the presence of three different external field strengths, that is, $30$ G, $90$ G and $150$ G respectively, demonstrating an increase in the droplet's deformation with an increasing magnetic field strength.

This behavior can be described by the Navier-Stokes equations in conjunction with the additional magnetic force (eq.~\ref{eq:interfacial_force}). We can thus calculate the droplet geometries as a function of magnetic field strength. Fig.~\ref{fig:2}b displays the droplet deformation in the presence of different magnetic fields pointing upwards along the z-axis, calculated using the finite element method. The model shows the dependence of the droplet's peak height on the strength of the magnetic field. In panel (c) of fig.~\ref{fig:2} we display the experimental realization of this "wobbling" motion, as the magnetic field rotates in the $x-z$ plane. The droplet wobbles at twice the frequency of the rotating field due to the squared force term in eq.~\ref{eq:interfacial_force}; that is, if the magnetization of the droplet $\boldsymbol{M}$ rotates at an angular frequency $\omega$, the magnetic force rotates at $2\omega$.

\begin{figure}[!ht]
    \centering
    \includegraphics[width = 0.95\linewidth]{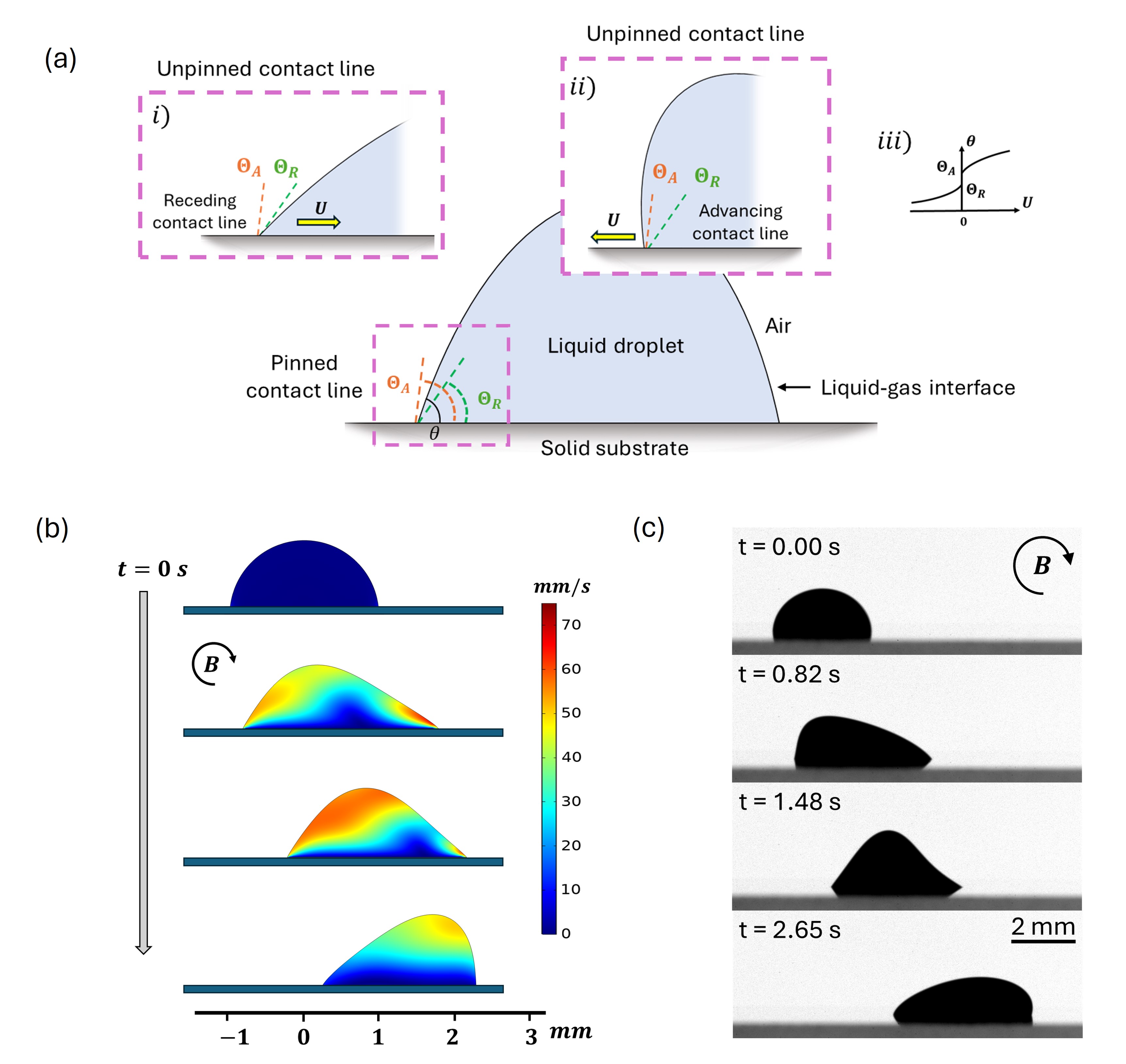}
    \caption{Migration of ferrofluid droplets in rotating fields. (a) A schematic representation of contact angle pinning. Insets ($i$) \& ($ii$) - representation of receding and advancing contact lines. ($iii$) - Depiction of contact line hysteresis diagram (contact line velocity $U$ vs. the dynamic contact angle $\theta$, adopted from \cite{dussan1979spreading}. (b) Finite element simulations demonstrating the migration of a 2D droplet due to the wobbling motion of fluid-air interface. The magnetic field is rotating clockwise at $10$ Hz in the $x-z$ plane. (c) Experimental photographs of a ferrofluid droplet undergoing the motion predicted by the finite element model. }
    \label{fig:3}
\end{figure}

The placement of the droplet on a solid substrate creates a pair of contact angles between the liquid-gas and liquid-solid interfaces. The mobility of the contact lines is a critical factor in determining the motion of the whole droplet. It is experimentally known that contact lines can become pinned whenever the dynamic contact angles $\theta(t)$ lie within a finite interval $\Theta_R < \theta(t) < \Theta_A$ of their static receding and advancing counterparts, (see fig.~\ref{fig:3}a). This phenomenon, known as contact angle hysteresis, occurs due to surface roughness, chemical contamination and other microscopic interactions of the surface with the fluid. The substrate used in the experiments (fig.~\ref{fig:2}c) has a contact angle hysteresis of roughly $15^\circ$, that is, $\Theta_A-\Theta_R \approx 15^\circ$. 
The wobbling motion of the droplet causes both contact angles of the droplet to oscillate (see supplementary video S1). This allows the pinned contact lines of the droplet to overcome contact angle hysteresis and become unpinned (cf. fig.~\ref{fig:3}a($i$)-($ii$)). 
The incipient motion of both contact lines over a full cycle of rotation of the magnetic field causes the droplet to become displaced relative to its initial position, thereby inducing migration.
The direction of the droplet migration follows the same `sense' as the magnetic field, that is, the droplet moves along the positive $x$-axis for a field rotating clockwise in the x-z plane, and vice-versa.
Using the finite element analysis model, we observe this property of ferrofluid droplets to migrate due to periodic wobbling of the interface  (see supplementary video S2). Fig.~\ref{fig:3}b shows the simulated motion of a 2-dimensional droplet via chronologically arranged snapshots. Here, the color bar shows the magnitude of fluid velocity in the interior of the droplet (see Methods section \ref{sec:method_theory} for material parameters used in the model). The model is in qualitative agreement with the motion observed in the laboratory  (see supplementary video S3). Fig. \ref{fig:3}c shows experiments where the migration of the droplet was induced using an external magnetic field of $100$ G rotating at $10 $ Hz.

\begin{figure}[!ht]
    \centering
    \includegraphics[width = 0.45\linewidth]{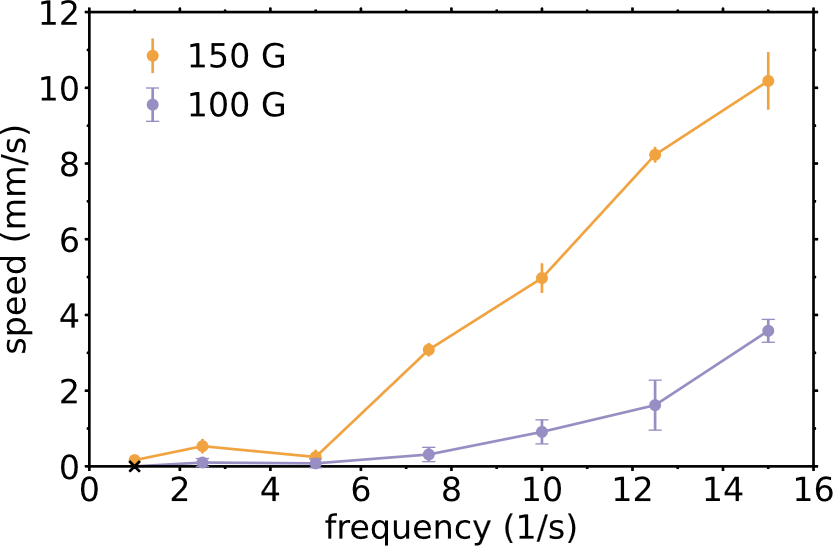}
    \caption{Ferrofluid migration velocity as a function of the rotating magnetic field frequency for two values of the field amplitude. }
    \label{fig:3-2}
\end{figure}

In the Stokes-flow regime (no inertia) and on a perfectly smooth substrate (no-contact angle hysteresis), during the first half of the deformation cycle, a wobbling droplet would deform into a sequence of geometric conformations that are mirror images (about $z$ axis) of the conformations during the second half. Therefore, under such conditions, the droplet should symmetrically move back and forth such that after every cycle, it returns back to the original position. This symmetry of the forward and backward motions is broken in our system due to the inertia of the fluid flow inside the droplet and the existence of contact angle hysteresis. 
The fluid flow depends on the amplitude and frequency of the magnetic field and therefore can be used to control the speed of the droplet motion.
Fig. \ref{fig:3-2} displays the droplet speed as a function of field frequency for two different values of field amplitudes, $100$ G and $150 $ G, respectively. Each data point in the plot is an average of five separate experiments using a new droplet on a clean glass substrate and the standard deviation of the measurements are plotted as error bars. In the experimental system, we observe a critical value of magnetic field and frequency below which the droplet's contact lines oscillate in place and the droplet does not exhibit locomotion. 
Even though the contact lines of this `immobile' configuration oscillate, the droplet does not move overall. One explanation for this phenomenon is the existence of surface inhomogeneities. This is confirmed by the different values taken by the contact angle hysteresis interval, measured at different positions on the glass substrate.  For small values of magnetic field amplitude and frequency, the droplet speeds are low and they cannot overcome hysteresis effects. However, we observe that beyond a critical field amplitude and frequency, the droplet starts to migrate on the solid substrate.




\begin{figure}[!ht]
    \centering
    \includegraphics[width = \linewidth]{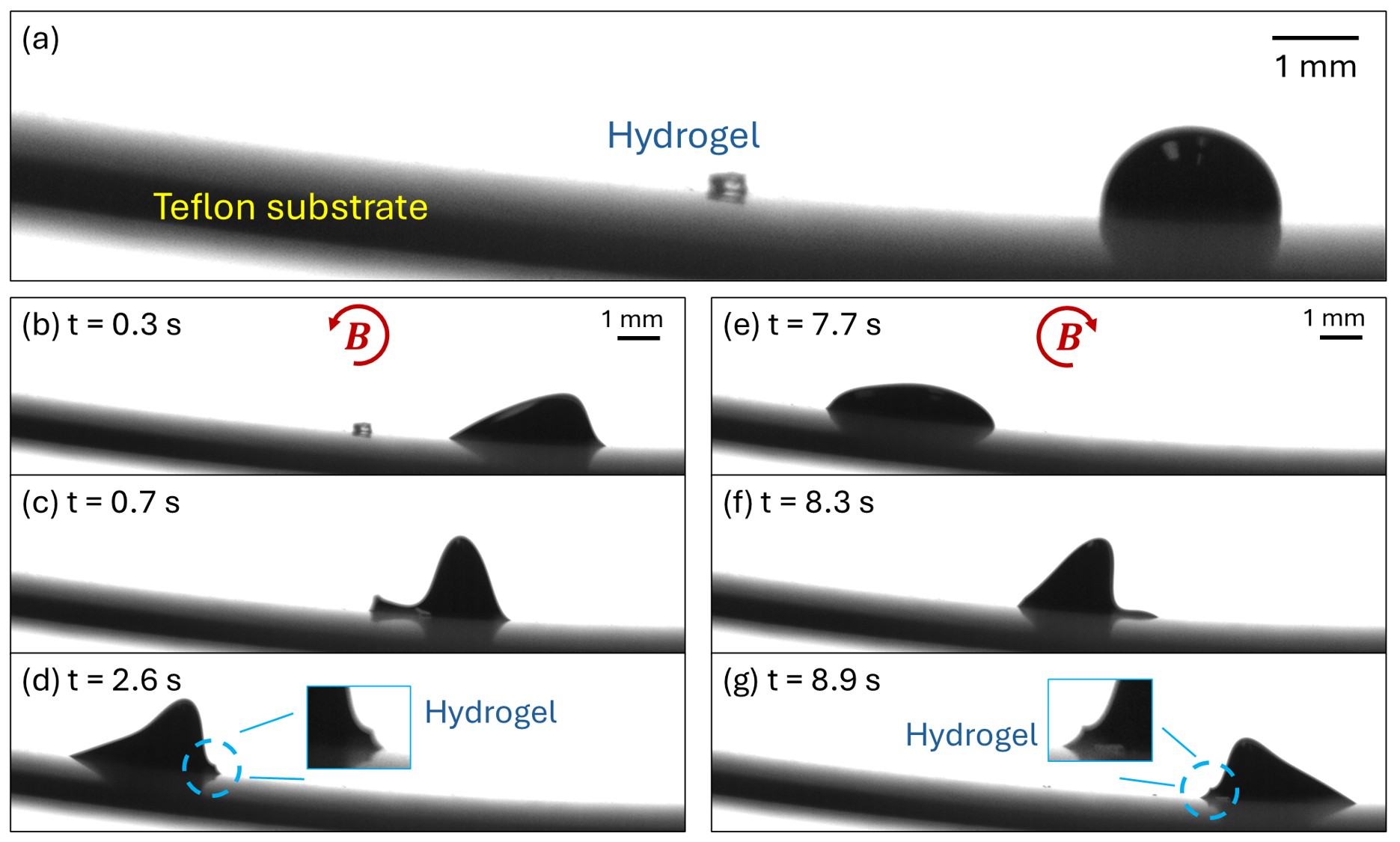}
    \caption{Experimental photographs of ferrofluid droplet picking up and transporting cargo. (a) Photograph showing the initial state of the droplet and cargo. (b)-(d) Chronologically arranged images of the ferrofluid droplet as it moves up an inclined plane to pick up the cargo. (e)-(g) Chronologically arranged images of the ferrofluid droplet as it moves down the inclined plane to deliver the cargo. }
    \label{fig:5}
\end{figure}

By controlling the axis of rotation of the magnetic field we can manoeuvre the droplets in any arbitrary direction on a solid substrate and in addition they can be induced to travel up or down inclined planes. The droplets can also interact with materials that lie along the path of their motion. This ability of the droplets can be used to clean surface impurities, harvest materials, and transport matter to desired locations on the substrate's surface. 
To demonstrate this functionality, we deposit a ferrofluid droplet at the bottom of a curved Teflon substrate, and place a small cube of a soft PEGMEA-based hydrogel (cargo) near the droplet (see fig.~\ref{fig:5}a). We drive the droplet uphill in the direction of the cargo using a magnetic field of $125$~G rotating counterclockwise at $10$~Hz until the droplet overruns the cargo (see figs. \ref{fig:5}b-d). This allows the droplet to pick up and move the cargo along with it. We then flip the direction of the field rotation to clockwise causing the droplet and cargo to move downhill together as shown in figs. \ref{fig:5}e-g (also see supplementary video S4). This shows our ability to control the motion of ferrofluid droplets on complex surfaces and also use them to transport matter. 

The model can also be used to analyse the motion of the droplet under different environmental conditions. 
For example, our numerical calculations show that for surfaces with large contact angle hysteresis ($\Theta_A - \Theta_R \sim 50^\circ$) and low surface tension, the droplet moves in the opposite `sense' relative to the magnetic field circulation, that is, droplets travel in negative $x$ direction for a field rotating clockwise in the $x-z$ plane (see supplementary video S5, paramaters used are given in methods section \ref{sec:method_theory}).  This reversed motion is associated with the inability of the right contact line to overcome the large hysteresis in the first half stroke of the clockwise rotating cycle. However, synchronization of the second half stroke with the fluid back-flow towards the left contact line, leads the latter to overcome hysteresis effects and sets the droplet into motion.  

\section{Conclusions}
In this article, we demonstrated the mechanism by which droplets of magnetic liquids can be manipulated over a solid substrate using rotating magnetic fields. 
We find that in the regime of negligible torques (no internal rotations), ferrofluid droplets can move via the deformation of its liquid-gas interface.
The rotation of the magnetic field creates periodic deformations in the droplet interface causing it to wobble. The wobbling interface creates fluid flows inside the droplet and inertia of the fluid leads the droplets to migrate. The speed of droplet motion is controlled by the amplitude and frequency of the magnetic field.
The droplets move in the same sense relative to the magnetic field rotation, that is, the droplets displace in the positive $x$ direction for a clockwise rotating field. 
We demonstrate this phenomenon of droplet motion through finite element modeling and experiments performed using water-based ferrofluid placed of a hydrophobic substrate. 

The finite element model allows us to explore different range of environmental and material parameters inaccessible to our experimental system. For instance, the model predicts that for substrates with large contact angle hysteresis, the droplets move in the opposite direction. Our work can also be extended to theoretically analyze more complex systems such as oil-based ferrofluid droplets in aqueous environments \cite{zakinyan2012motion} as well as to explore functions such as cleaning surface impurity, interacting with complex geometry like curved surfaces, dragging matter, and delivering cargo.



\section{Acknowledgement}
This work was funded by the University of Chicago Materials Research Science and Engineering Center, which is funded by National Science Foundation under award number DMR-2011854.
We thank Chloe Lindeman at University of Chicago for helping us to use the Kr\"{u}ss drop shape analyzer (DSA100).
This work made use of the shared facilities at the University of Chicago Materials Research Science and Engineering Center, supported by National Science Foundation under award number DMR-2011854.

\section{Author Contribution}
A.A. and S-Y.C. contributed equally. 
A.A. developed the theoretical framework and designed the finite element analysis model. S-Y.C. carried out the experiments.  
A.A., E.K., S-Y.C., M.M.D and M.O.d.l.C. analyzed the data. A.A., S-Y.C., M.O.d.l.C. and E.K. wrote the manuscript.
M.I.K and B.F. synthesised the hydrophobic glass plates used in the experiments.
M.M.D and M.O.d.l.C directed the research.

\section{Methods}
\subsection{Experimental method} \label{sec:method_exp}

We use commercial ferrofluid (FerroTec, EMG 700) and quantify the contact angle hysteresis of the ferrofluid on the hydrophobic surface with a drop shape analyzer (Kr\"{u}ss, DSA100). We first deposit a ferrofluid drop on a clean, hydrophobic slide while the other side of the drop remains attached to a synringe needle. We then gradually increase the drop volume ($0.5 \ \mu$L$/$s) and measure the advancing contact angle when the contact line is sliding both along the substrate and the needle. We then decrease the drop volume ($-0.5 \ \mu$L$/$s) and measure the receding contact angle. The reported static advancing angle is $90^{\circ}$ and the receding angle $75^{\circ}$.
We further check the inhomogeneity of the substrate by depositing ferrofluid drops on various locations of the substrate, and apply a DC field to stretch the drop against the gravity. As the drop deforms, we observe that the drop develops a strong pinned contact line at some locations while not at other locations. To minimize the pinned contact line problem, we only perform experiments at locations with no surface inhomogeneity, e.g. where we observe no pinned contact line. (All details on the instruments and materials used are given in the SI).

\subsection{Computational method} \label{sec:method_theory}
To solve the continuum model we used a commercial finite analysis software, COMSOL \cite{comsol2024comsol}, and utilized COMSOL's AC/DC module to solve for the magnetic fields and the computational fluid dynamics module to solve the Navier-Stokes equations with magnetic force applied at the droplet interface. The droplet is modeled using the \textit{moving mesh} method defined within the framework of Arbitrary Lagrangian-Eulerian (ALE) formulation \cite{donea2004arbitrary}, where the fluid-air interface is described as a geometric surface and the interfacial forces are directly applied on the boundary of the fluid domain. The advantage of using the moving mesh method is that it provides a very sharp and accurate interface.

Here, we consider the case of very small relaxation time of the magnetization, which is thus collinear with the magnetic field (quasi-stationary theory, cf. \cite{rosensweig2013ferrohydrodynamics}). In this case the magnetization is described by the Langevin function 
\begin{equation}
    \label{eq:magnetization}
    M(H) = M_s \left( \coth(\tau H) - \frac{1}{\tau H}\right), \quad \tau = \frac{m_d}{k_BT},
\end{equation}
where, $M_s = nm_d$ is the saturation magnetization, 
$m_d = M_d V$ is the magnetic moment of a single subdomain particle, $V$ is the particle volume, $M_d$ is the domain magnetization of dispersed ferromagnetic material and $n$ is the number density of the magnetic grains.
The magnetic force is then described by eq. \ref{eq:interfacial_force}.





We implemented contact angle hysteresis using the approach developed by Cai and Song \cite{cai2021implementing}. On a perfectly smooth surface with no contact angle hysteresis, the liquid-gas interface of a viscous droplet at rest makes an angle $\theta_{eq}$ with the surface. This angle is determined by the balance of interfacial forces given by the 
Young-Laplace equation \cite{deGennes1985}. If the dynamic contact angle of the droplet differs from $\theta_{eq}$, the contact line moves until $\theta$ and $\theta_{eq}$ are equal. In order to implement contact angle hysteresis, $\theta_{eq}$ is changed depending upon whether the contact line is in a \textit{pinned state} or an \textit{unpinned state} \cite{cai2021implementing}. In the \textit{pinned state} where $\Theta_R \leq \theta \leq \Theta_A$, the energy of the liquid-solid interface is constantly adjusted such that, $\theta_{eq} = \theta$. This causes the contact lines to become stationary. 
On the other hand, when the dynamic contact angle exceeds the static advancing angle, that is, $\theta>\Theta_A$, we fix $\theta_{eq} = \Theta_A$. In this case, since the dynamic contact angle exceeds the equilibrium contact angle, the contact lines are free to move. Similarly, when $\theta < \Theta_R$, we fix $\theta_{eq} = \Theta_R$.

The material parameters used in the model are fluid density $\rho = 1290\textrm{ kg/m}^3$, viscosity $\mu = 5$~cP, surface tension $\gamma= 51$~mN/m, saturation magnetization $M_s = 355$ G, receding static contact angle $\Theta_R = 75^\circ$, advancing static contact angle $\Theta_A = 90^\circ$. 

Parameters used in supplementary video S5 are fluid density $\rho = 1290\textrm{ kg/m}^3$, viscosity $\mu = 10$~cP, surface tension $\gamma= 30$~mN/m, saturation magnetization $M_s = 355$ G, receding static contact angle $\Theta_R = 55^\circ$, advancing static contact angle $\Theta_A = 110^\circ$.


\bibliographystyle{unsrt} 
\bibliography{SYCref,bib_theory} 

\begin{thebibliography}{10}

\bibitem{aggarwal2022controlling}
Aaveg Aggarwal, Chuang Li, Samuel~I Stupp, and Monica~Olvera de~la Cruz.
\newblock Controlling the shape morphology of origami-inspired photoresponsive
  hydrogels.
\newblock {\em Soft matter}, 18(11):2193--2202, 2022.

\bibitem{zhao2019soft}
Yusen Zhao, Chen Xuan, Xiaoshi Qian, Yousif Alsaid, Mutian Hua, Lihua Jin, and
  Ximin He.
\newblock Soft phototactic swimmer based on self-sustained hydrogel oscillator.
\newblock {\em Science Robotics}, 4(33):eaax7112, 2019.

\bibitem{palacci2013photoactivated}
J{\'e}r{\'e}mie Palacci, Stefano Sacanna, Adrian Vatchinsky, Paul~M Chaikin,
  and David~J Pine.
\newblock Photoactivated colloidal dockers for cargo transportation.
\newblock {\em Journal of the American Chemical Society}, 135(43):15978--15981,
  2013.

\bibitem{driscoll2019leveraging}
Michelle Driscoll and Blaise Delmotte.
\newblock Leveraging collective effects in externally driven colloidal
  suspensions: Experiments and simulations.
\newblock {\em Current opinion in colloid \& interface science}, 40:42--57,
  2019.

\bibitem{jackson2017ionic}
Brandon~A Jackson, Kurt~J Terhune, and Lyon~B King.
\newblock Ionic liquid ferrofluid interface deformation and spray onset under
  electric and magnetic stresses.
\newblock {\em Physics of Fluids}, 29(6), 2017.

\bibitem{zhang2022polar}
Bo~Zhang, Hang Yuan, Andrey Sokolov, Monica~Olvera de~la Cruz, and Alexey
  Snezhko.
\newblock Polar state reversal in active fluids.
\newblock {\em Nature Physics}, 18(2):154--159, 2022.

\bibitem{pradillo2019quincke}
Gerardo~E Pradillo, Hamid Karani, and Petia~M Vlahovska.
\newblock Quincke rotor dynamics in confinement: rolling and hovering.
\newblock {\em Soft Matter}, 15(32):6564--6570, 2019.

\bibitem{li2021chemically}
Siyu Li, Daniel~A Matoz-Fernandez, Aaveg Aggarwal, and Monica Olvera de~la
  Cruz.
\newblock Chemically controlled pattern formation in self-oscillating elastic
  shells.
\newblock {\em Proceedings of the National Academy of Sciences},
  118(10):e2025717118, 2021.

\bibitem{maeda2007self}
Shingo Maeda, Yusuke Hara, Takamasa Sakai, Ryo Yoshida, and Shuji Hashimoto.
\newblock Self-walking gel.
\newblock {\em Advanced Materials}, 19(21):3480--3484, 2007.

\bibitem{manna2023chemically}
Raj~Kumar Manna, Oleg~E Shklyaev, and Anna~C Balazs.
\newblock Chemically driven multimodal locomotion of active, flexible sheets.
\newblock {\em Langmuir}, 39(2):780--789, 2023.

\bibitem{sarkhosh_manipulation_2023}
Mohammad~Hosein Sarkhosh, Masoud Yousefi, Mohamad~Ali Bijarchi, Hossein
  Nejat~Pishkenari, and Kimia Forghani.
\newblock Manipulation of ferrofluid marbles and droplets using repulsive force
  in magnetic digital microfluidics.
\newblock {\em Sensors and Actuators A: Physical}, 363:114733, December 2023.

\bibitem{nguyen_kinematics_2007}
Nam-Trung Nguyen, Ali Beyzavi, Kon~Meng Ng, and Xiaoyang Huang.
\newblock Kinematics and deformation of ferrofluid droplets under magnetic
  actuation.
\newblock {\em Microfluidics and Nanofluidics}, 3(5):571--579, October 2007.

\bibitem{nguyen_magnetowetting_2010}
Nam-Trung Nguyen, Guiping Zhu, Yong-Chin Chua, Vinh-Nguyen Phan, and Say-Hwa
  Tan.
\newblock Magnetowetting and {Sliding} {Motion} of a {Sessile} {Ferrofluid}
  {Droplet} in the {Presence} of a {Permanent} {Magnet}.
\newblock {\em Langmuir}, 26(15):12553--12559, August 2010.
\newblock Publisher: American Chemical Society.

\bibitem{fan_reconfigurable_2020}
Xinjian Fan, Xiaoguang Dong, Alp~C. Karacakol, Hui Xie, and Metin Sitti.
\newblock Reconfigurable multifunctional ferrofluid droplet robots.
\newblock {\em Proceedings of the National Academy of Sciences},
  117(45):27916--27926, November 2020.
\newblock Publisher: Proceedings of the National Academy of Sciences.

\bibitem{li2020fast}
Chuang Li, Garrett~C Lau, Hang Yuan, Aaveg Aggarwal, Victor~Lopez Dominguez,
  Shuangping Liu, Hiroaki Sai, Liam~C Palmer, Nicholas~A Sather, Tyler~J
  Pearson, et~al.
\newblock Fast and programmable locomotion of hydrogel-metal hybrids under
  light and magnetic fields.
\newblock {\em Science robotics}, 5(49):eabb9822, 2020.

\bibitem{li2023magnetic}
Meng Li, Aniket Pal, Junghwan Byun, Gaurav Gardi, and Metin Sitti.
\newblock Magnetic putty as a reconfigurable, recyclable, and accessible soft
  robotic material.
\newblock {\em Advanced Materials}, 35(48):2304825, 2023.

\bibitem{Kirkinis2017}
E.~Kirkinis.
\newblock Magnetic torque-induced suppression of van-der-{W}aals-driven thin
  liquid film rupture.
\newblock {\em Journal of Fluid Mechanics}, 813:991--1006, 2017.

\bibitem{aggarwal2023activity}
A~Aggarwal, E~Kirkinis, and M~Olvera De~La~Cruz.
\newblock Activity-induced migration of viscous droplets on a solid substrate.
\newblock {\em Journal of Fluid Mechanics}, 955:A10, 2023.

\bibitem{sun_exploiting_2022}
Mengmeng Sun, Bo~Hao, Shihao Yang, Xin Wang, Carmel Majidi, and Li~Zhang.
\newblock Exploiting ferrofluidic wetting for miniature soft machines.
\newblock {\em Nature Communications}, 13(1):7919, December 2022.
\newblock Publisher: Nature Publishing Group.

\bibitem{fan_ferrofluid_2020}
Xinjian Fan, Mengmeng Sun, Lining Sun, and Hui Xie.
\newblock Ferrofluid {Droplets} as {Liquid} {Microrobots} with {Multiple}
  {Deformabilities}.
\newblock {\em Advanced Functional Materials}, 30(24):2000138, 2020.
\newblock \_eprint:
  https://onlinelibrary.wiley.com/doi/pdf/10.1002/adfm.202000138.

\bibitem{katsikis_synchronous_2015}
Georgios Katsikis, James~S. Cybulski, and Manu Prakash.
\newblock Synchronous universal droplet logic and control.
\newblock {\em Nature Physics}, 11(7):588--596, July 2015.
\newblock Publisher: Nature Publishing Group.

\bibitem{chaves2008spin}
Arlex Chaves, Markus Zahn, and Carlos Rinaldi.
\newblock Spin-up flow of ferrofluids: Asymptotic theory and experimental
  measurements.
\newblock {\em Physics of Fluids}, 20(5), 2008.

\bibitem{rosensweig1990magnetic}
RE~Rosensweig, J~Popplewell, and RJ~Johnston.
\newblock Magnetic fluid motion in rotating field.
\newblock {\em Journal of Magnetism and Magnetic Materials}, 85(1-3):171--180,
  1990.

\bibitem{rosensweig2013ferrohydrodynamics}
Ronald~E Rosensweig.
\newblock {\em Ferrohydrodynamics}.
\newblock Courier Corporation, 2013.

\bibitem{dussan1979spreading}
EB~Dussan.
\newblock On the spreading of liquids on solid surfaces: static and dynamic
  contact lines.
\newblock {\em Annual Review of Fluid Mechanics}, 11(1):371--400, 1979.

\bibitem{zakinyan2012motion}
Arthur Zakinyan, Oksana Nechaeva, and Yu~Dikansky.
\newblock Motion of a deformable drop of magnetic fluid on a solid surface in a
  rotating magnetic field.
\newblock {\em Experimental Thermal and Fluid Science}, 39:265--268, 2012.

\bibitem{comsol2024comsol}
{COMSOL Multiphysics\textsuperscript{\tiny\textregistered} v. 6.1 }.
\newblock {www.comsol.com.}

\bibitem{donea2004arbitrary}
Jean Donea, Antonio Huerta, J-Ph Ponthot, and Antonio Rodr{\'\i}guez-Ferran.
\newblock Arbitrary l agrangian--e ulerian methods.
\newblock {\em Encyclopedia of computational mechanics}, 2004.

\bibitem{cai2021implementing}
Zheren Cai and Yanlin Song.
\newblock Implementing contact angle hysteresis in moving mesh-based two-phase
  flow numerical simulations.
\newblock {\em ACS omega}, 6(51):35711--35717, 2021.

\bibitem{deGennes1985}
P.G. de~Gennes.
\newblock Wetting: statics and dynamics.
\newblock {\em {Reviews of Modern Physics}}, 57(3):827, 1985.

\end{thebibliography}

\newpage

\section*{Supplemental Information: Wobbling and Migrating Ferrofluid Droplets}

\section*{Supporting videos}

Supporting videos can be viewed at: \url{https://doi.org/10.6084/m9.figshare.26023099.v1}
\begin{itemize}
    \item \textbf{Video S1: Periodic oscillations of droplet contact lines}.  
The wobbling liquid-gas interface creates periodic oscillations in the contact angles. The field amplitude and frequency used here is $25$ G and $2.5$ Hz respectively. The orange dashed line shows the liquid-solid interface. The two solid orange lines depict the slope of the liquid-gas interface at the contact points. 

\item \textbf{Video S2: Finite element analysis of droplet migration}.  
Finite element analysis simulation demonstrating the motion of a 2D droplet place of a solid substrate in the presence of a rotating magnetic field. The droplet moves in the positive $x$ direction and magnetic field rotates clockwise.

\item \textbf{Video S3: Migrating droplet}.  
A ferrofluid droplet migrating on a solid substrate due to periodic deformation of the liquid-gas interface induced by a rotating magnetic field. The droplet moves in the positive $x$ direction and magnetic field rotates clockwise. The field amplitude and frequency are $150$ G and $10$ Hz respectively.

\item \textbf{Video S4: Cargo pickup and transportation}.  
Ferrofluid drop deposited on a curved Teflon surface. The droplet is being controlled to pick up a piece of hydrogel placed on the left side of the drop. The drop climbs up the curved surface and picks up the hydrogel as the field rotates counterclockwise. Then in the presence of clockwise rotating field, the droplet climbs down the surface, dragging the hydrogel with it.
 
\item \textbf{Video S5: Finite element analysis of a droplet moving in the opposite direction}.  
Finite element analysis simulation predicting backward motion of a 2D droplet due to large contact angle hysteresis. The droplet moves in the negative $x$ direction and magnetic field rotates clockwise. 

\end{itemize}

\section*{Experimental methods}

We connect two bipolar amplifiers (Kepco BOP 50-8ML) to two pairs of Helmholtz coils. We send two sinusoidal signals through a DAC (Measurement computing USB-1208HS-4AO) to control the phase and the value of the output current. 
We use a magnetic field probe (Devonian, TD8620) and apply a DC current to measure the magnetic field strength. We further use a Hall-effect magnetic sensor (Sentron, 2SA-10) to verify that the field strength remains the same under an AC field.
We use a high speed camera (Phantom, VEO 640) to capture the deformation and transportation of ferrofluid drops. The camera is slightly tilted ($0.8^{\circ}$) to capture the reflection of the ferrofluid drops, which helps the image analysis process. 

For the hydrophobic substrate, we first clean glass slides by sequential rinsing with acetone and isopropyl alcohol (IPA), followed by deionized (DI) water. The slides are then air-dried. The slippery hydrophobic surface coating is fabricated through a one-step self-catalyzed polymer brush grafting method. This involves anchoring chlorine terminated PDMS polymers with various molecular weights (MW 425-650 and MW 2000-4000) onto the glass substrates. The substrates are first treated with air plasma for 20 minutes to functionalize the surface with hydroxyl groups. The plasma treated substrates are then inverted and placed on the cover of a Petri dish containing 800 µL of the chlorine terminated PDMS oligomer. The assembly is subjected to a 60-minute treatment in a vacuum oven at $60^{\circ}$ (Yamato Vacuum drying oven). For post-treatment, the substrates are immersed in a toluene bath, and subsequently rinsed with DI water and air-dried. 
Before each ferrofluid experiment, we rinse the slides in MillQ water, then sonicate the slides in isopropyl alcohol for $10$~minutes to clean them, and finish by drying them with nitrogen. 

To control the size of the ferrofluid drop, we use a $10~\mu L$ pipette and only withdraw $7.5~\mu L$ amount of the liquid. When depositing, we slowly and steadily push the liquid out so it naturally falls to the substrate due to gravity. 
For each experiment, we apply the magnetic field right after depositing the drop on the substrate, which minimizes the influence of evaporation. After every single experiment, we remove the drop and clean the residual using MilliQ water, and then sonicate the slides in isopropyl alcohol for $10$~minutes. Every reported datum comes from a fresh drop to erase the possibility of magnetic hysteresis. 

\end{document}